\documentclass{article}
\usepackage[utf8]{inputenc}
\usepackage{spconf,amsmath,amssymb,graphicx,hyperref}
\usepackage{booktabs,array,multirow,xcolor,colortbl}
\usepackage{tikz}
\usepackage{pgfplots}
\usepackage{arydshln}
\usepackage{enumitem}

\usepackage{orcidlink}
\newcommand{\orcid}[1]{%
  \begingroup
  \hypersetup{pdfborder={0 0 0}}%
  \,\textnormal{\orcidlink{#1}}%
  \endgroup
}

\pgfplotsset{compat=1.14}
\usetikzlibrary{arrows.meta,positioning,fit,backgrounds,decorations.pathreplacing}

\definecolor{cMerge}{HTML}{1B6CA8}
\definecolor{cBase}{HTML}{B0562B}
\definecolor{cAccent}{HTML}{4E7A51}
\definecolor{cGrey}{HTML}{6B6B6B}
\definecolor{hlrow}{HTML}{EAF2F8}

\newcommand{\dpair}{\ensuremath{\Delta_{\mathrm{pair}}}}
\newcommand{\dlabel}{\ensuremath{\Delta_{\mathrm{label}}}}
\newcommand{\wys}{\ensuremath{W_{y,s}}}

\title{Correspondence Is Not Transfer: A Pairing-Controlled Study of Clinical Audio--Metadata Alignment}
\name{\shortstack{
Liu He$^1$\orcid{0000-0002-6128-4292},
Xi Wang$^2$\orcid{0009-0004-6175-3986},
Jiahong Yuan$^1$\orcid{0009-0008-2162-7167},
Yuanchao Li$^{*3}$\orcid{0000-0003-4266-2005}\thanks{$^*$Corresponding author.}
}}
\address{$^1$University of Science and Technology of China, China \\ \quad $^2$Independent Researcher\quad $^3$University of Cambridge, UK \\
\small \texttt{heliummn@mail.ustc.edu.cn},\quad \texttt{jiahongyuan@ustc.edu.cn},\quad \texttt{yl809@cam.ac.uk}}

\begin{document}
\ninept
\maketitle
\begin{abstract}
Clinical audio models are often trained by aligning recordings with
participant metadata or text derived from it. This metadata is rarely a clean
disease signal, since a single profile also contains symptoms, age, sex,
missing responses, and recruitment information. We examine what alignment adds
to frozen audio features by varying only the audio--metadata pairing. We keep
the audio backbones fixed and train only a projector under correct pairing and
three shuffled conditions that progressively preserve the disease label and
recorded sex. We then evaluate each projector on profile retrieval, attribute
probing, and disease classification on covariate-matched test sets. On
UKCOVID, correct pairing clearly improves retrieval and makes sex easier to
decode, confirming that the pairing manipulation works. Most of the sex-probe
gap disappears when the control also preserves recorded sex, while a small
retrieval gap remains. Matched disease AUROC differences are small, and every
interval includes zero, so modest benefits or harms are not ruled out. The
same pattern appears when we rerun the frozen pipeline on CODA TB, Cambridge
COVID-19 Sounds, and Coswara, where retrieval and sex-probe contrasts are
consistently positive but disease contrasts are not. We find no consistent
evidence that better profile correspondence translates into better disease
prediction and suggest reporting these outcomes separately.
\end{abstract}


\section{Introduction}
\label{sec:intro}

Cough recordings in screening studies usually come with metadata. These
records include symptom checklists, age brackets, recorded sex, recruitment
channels, and patterns of unanswered questions. Because diagnostic labels are
scarce, recent clinical audio work has begun using this information as a
training signal, either as structured metadata or as short texts generated
from it
\cite{soni2022contrastive,moummad2023metadata,kim2024bts,siam2026respiramfm}.
Audio-language systems make this approach explicit by contrastively aligning
audio representations with participant profiles
\cite{ilerisoy2026zeroshot,ando2025slap,min2026semantic}. The expectation is
that a model that can recover the clinical context of a recording has also
learned something about the disease. Whether that expectation holds is difficult to assess. A contrastive loss
rewards any feature that helps identify the assigned profile. Profiles can be
identifiable for reasons unrelated to disease, such as unusual symptom
combinations, rare age--sex pairs, or missingness patterns typical of a
recruitment site. The representation may therefore encode disease information,
a demographic shortcut, or a fingerprint of the metadata form, while the
training curve looks the same in all three cases \cite{arasteh2026linkage}.
Prior findings give specific reasons for concern. COVID screening performance
drops after adjustment for symptoms and demographics \cite{coppock2024audio},
alignment helps or hurts depending on what the two modalities share
\cite{fang2026align}, and contrastive objectives settle on shortcuts whenever
one is available \cite{robinson2021shortcut}.

To separate these possibilities, we vary only the audio--metadata pairing during training and hold everything else fixed. A global shuffle is
a natural baseline, but it changes both the link to the original participant
and the disease-label mix of the metadata assigned to each recording. We
therefore compare four pairing schemes. Correct ($C$) retains the observed
pair. Within-label ($W$) assigns metadata from a different participant with
the same disease label. Global ($G$) shuffles without label or sex
restrictions. The fourth arm, \wys{}, also preserves recorded sex. Disease
labels are used to construct these controls but do not appear in the metadata
text. The $C-W$ comparison tests what changes when exact pairing is replaced
by within-label shuffling, while $W-G$ tests the effect of restricting
metadata donors by disease label. Each trained projector is then frozen and evaluated on retrieval of the
paired profile, decodability of demographic and cohort attributes, and disease
prediction on test sets balanced for measured covariates. We interpret probe
accuracy as indicating how accessible an attribute is to a simple classifier,
not as evidence that a disease classifier uses it
\cite{ravichander2021probing}. Here, ``transfer'' refers to evaluation on a
matched split within a dataset, not to transferring a model across datasets.

\begin{figure*}[t]
\centering
\includegraphics[width=0.80\textwidth]{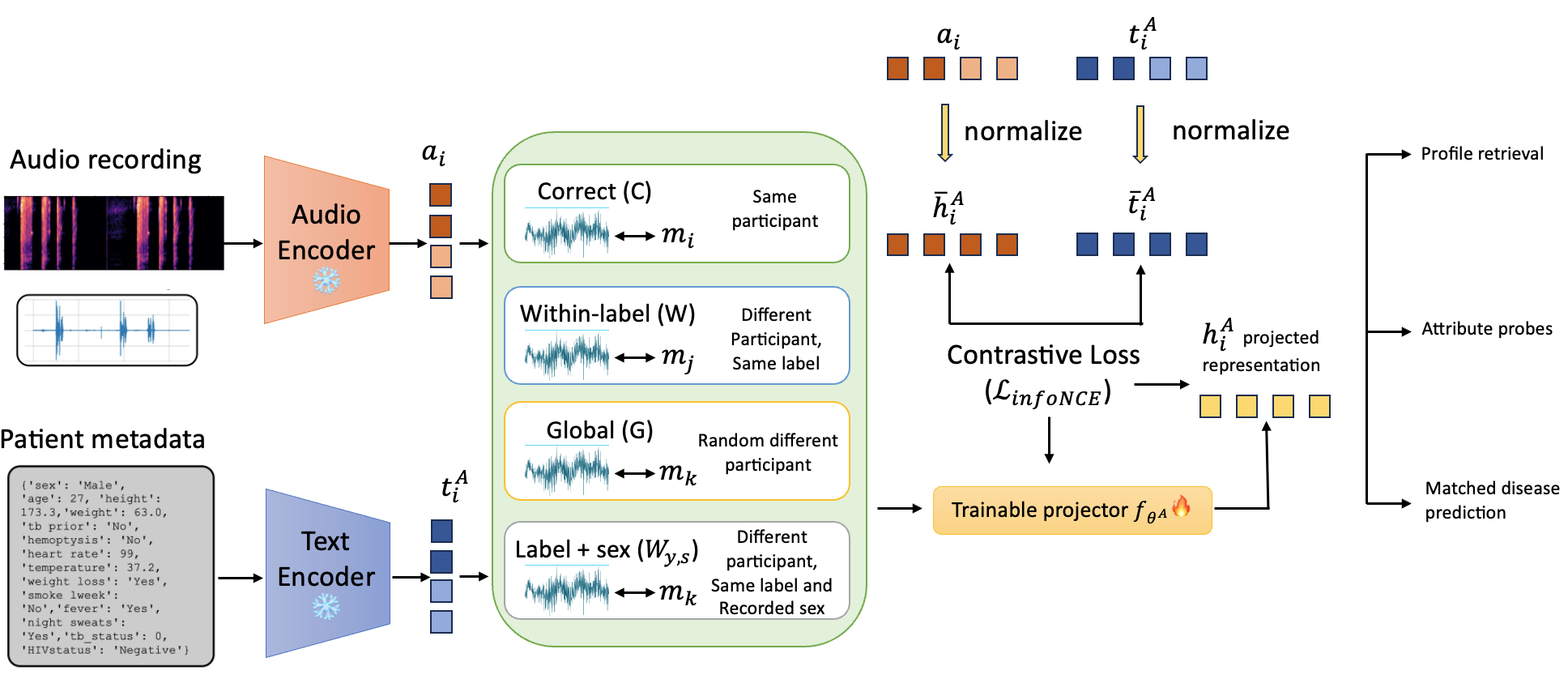}
\caption{Study design. The only intervention is which participant's metadata
is paired with each recording during alignment training. The frozen projectors
are then evaluated on profile retrieval, attribute probing, and disease
classification on matched test sets.}
\label{fig:design}
\end{figure*}

We conduct the study in two stages. UKCOVID is the discovery dataset used to
fix the pipeline, including the pairing arms, backbones, matching variables,
probes, calibration, bootstrap, the primary contrast ($C-\wys{}$), and the
primary metric. We then run CODA TB, Cambridge COVID-19 Sounds, and Coswara
with the frozen pipeline, without changing analysis settings after seeing the
external results. Figure~\ref{fig:design} summarizes the design. We find three
things. First, label-preserving shuffling provides a different training signal
from global shuffling, so a two-arm correct-versus-shuffled ablation mixes two
effects. Second, correct pairing reliably improves profile retrieval and sex
decodability, with most of the sex effect already present when the control
preserves recorded sex. Third, we find no consistent change in matched disease
AUROC, and across 12 dataset--backbone settings no interval excludes zero in
either direction. The paper's contribution is a fixed-pipeline separation of
profile correspondence, attribute decodability, and disease transfer.

\vspace{-5pt}

\section{Data and Methods}
\label{sec:method}

\subsection{Discovery dataset and external panel}

UKCOVID \cite{budd2024ukcovid} is the discovery dataset. Following
audio-quality filtering, the Standard training split contains 20,714
participants. We evaluate on the Standard test set ($n=11{,}119$), a
covariate-matched test set ($n=1{,}814$), and a larger participant-disjoint
matched set ($n=4{,}196$). Matching is particularly relevant for this cohort,
because recruitment source alone predicts COVID status with an AUROC of 99.66
on Standard train and 50.00 after matching. This association motivates matched
evaluation but does not itself show that audio classifiers exploit
recruitment-related information. UKCOVID serves both as the discovery dataset and as the basis for fixing the
external analysis pipeline. Before inspecting any external results, we fixed
the arm definitions, backbones, matching variables, probes, calibration,
bootstrap, primary contrast, primary metric, five seeds, and the use of the
500-epoch checkpoint. The external analyses use the archived configuration
without modification.

The external panel consists of CODA TB (1,081 eligible participants)
\cite{huddart2024coda}, Cambridge COVID-19 Sounds (989)
\cite{xia2021covidsounds}, and Coswara (1,701) \cite{bhattacharya2023coswara}.
Each dataset has a matched test set of 200 participants and uses the same
pairing arms, backbones, probes, calibration, bootstrap, primary contrast, and
metric. Three qualifications apply. CODA uses a match-first TB analysis,
Cambridge is a strict-COVID reconstruction of the Task-2 cough subset rather
than the official benchmark, and Coswara is matched before any model is run.
These qualifications limit generalizability, but the external analyses still
follow the unchanged protocol. Matching balances only the selected measured
covariates \cite{stuart2010matching}. We release cohort hashes and
protocol-lock logs, but not participant identifiers, pair lists, embeddings,
or predictions.
\vspace{-10pt}
\subsection{Pairing-controlled training}

All pairing conditions use frozen audio and text encoders. AST-6L retains the
first six blocks of an AudioSet-finetuned AST \cite{gong2021ast}; this depth
was chosen from respiratory-audio comparisons that did not use UKCOVID matched
performance. OPERA-CT is the released respiratory contrastive Transformer
\cite{zhang2024opera}; because its pretraining data include UKCOVID training
audio, we treat it as a domain-pretrained comparator rather than a clean
backbone. HeAR \cite{baur2024hear} is the third health-acoustic backbone. Each
backbone produces one pooled audio vector per participant. We keep the text representation simple by design. A frozen Phi-2 encoder
\cite{javaheripi2023phi2} embeds deterministic key--value schema strings with
a fixed field order and explicit missing values, and we do not generate free-form
clinical narratives. This makes the text target fully reproducible and the
included fields explicit. Disease labels, participant identifiers, recruitment
source, site, country, platform, device, timestamps, and recording artefacts
are all excluded from the text. Let $a_i,m_i,y_i,s_i$ denote the audio, metadata, disease label, and recorded
sex of participant $i$. For arm $A$, $\pi_A(i)$ identifies the participant whose
metadata is paired with audio $a_i$:
\begin{equation}
\begin{aligned}
 C &: \pi_C(i)=i,\\
 W &: \pi_W(i)\ne i,\quad y_{\pi_W(i)}=y_i,\\
 G &: \pi_G(i)\ne i,\\
 \wys{} &: \pi_{W_{y,s}}(i)\ne i,\quad
          (y_{\pi_{W_{y,s}}(i)},s_{\pi_{W_{y,s}}(i)})=(y_i,s_i).
\end{aligned}
\label{eq:pairing}
\end{equation}
Global shuffling imposes no label or sex restrictions, so some pairs share
a label or sex by chance. Across arms, the audio examples, initialization, optimizer, seeds,
batch order, dropout streams and training length are identical, and only $\pi_A$
changes. We write \dpair{} for $C-W$ and \dlabel{} for $W-G$, computed
separately for each evaluation. These are contrasts between pairing
constraints. They are not causal estimands, and they do not decompose the
representation into disjoint information sources. $R$ denotes the frozen
raw-audio reference. The \wys{} arm uses the same 20,714 UKCOVID training participants and rejects
self-pairs, singleton strata, and cross-stratum fallback. Because the schema
text contains only a small number of fields, different participants can have
identical profiles, although fewer than 7\% of shuffled pairs have the same
schema string. The contrasts serve different purposes. $C-W$ is the manipulation check and
the decodability contrast. It tests whether exact pairing changes retrieval or
attribute accessibility beyond what label-matched pairing already provides.
The prespecified clinical contrast, $C-\wys{}$, tests whether exact pairing
improves matched disease AUROC when the control also preserves recorded-sex
consistency. We report $C-W$ for retrieval and sex probing, and $C-\wys{}$ for
disease prediction.

\subsection{Training and evaluation}

Only an audio-side projector is trained. Let
$t_i^A=e_{\mathrm{text}}(m_{\pi_A(i)})$ be the frozen text embedding assigned
to participant $i$ under arm $A$. For a batch $\mathcal B$ of size $B$ we
minimize a one-directional InfoNCE loss \cite{oord2018cpc,radford2021clip}
over in-batch pairs:
\begin{equation}
\mathcal L_A=-\frac{1}{B}\sum_{i\in\mathcal B}
\log\frac{\exp(\bar f_\theta(a_i)^\top\bar t_i^A/\tau)}
{\sum_{j\in\mathcal B}\exp(\bar f_\theta(a_i)^\top\bar t_j^A/\tau)},
\quad \tau=0.07,
\label{eq:loss}
\end{equation}
where bars denote $\ell_2$ normalization. We train a SimCLR-style projector \cite{chen2020simclr} for 500 epochs
with five seeds. The final checkpoint is used without epoch selection and then
frozen for all evaluations. This reproduces the Stage-1 alignment of RespiraMFM, not
its full downstream Phi-2/LoRA system.

\textbf{Profile retrieval.} We use source-validation audio as queries to
retrieve metadata profiles from a candidate pool. Identical schema strings are
collapsed into a single target, and we report macro-averaged mean reciprocal
rank (MRR) over profiles so that frequent profiles do not dominate. As a
sensitivity check, we also restrict the candidate pool to profiles with the
same recorded sex or with the same label and sex. This changes the retrieval
task, not the representation.

\textbf{Attribute probing.} Regularized linear probes with identical
hyperparameters predict recorded sex, age, symptoms, cohort variables, and
disease from each frozen representation. Probe performance measures how
accessible an attribute is to that probe, not whether a disease classifier
uses it.

\textbf{Disease classification on matched test sets.} We report
AUROC~\cite{fawcett2006roc}, negative log-likelihood (NLL), and Brier score.
In the primary source-only evaluation, we select the classifier and calibrate
it on source data \cite{platt1999probabilistic} before evaluating it on the
matched target. Matched splits are not used to select either the projector or
the classifier. Reference models use frozen raw audio or metadata alone. A
fixed MLP serves as an additional nonlinear classifier. Two further
diagnostics are not source-only. A target-assisted variant trains, selects,
and calibrates on the larger matched set before evaluation on
participant-disjoint matched participants. A separate one-dimensional
recalibration tests whether the differences are confined to the probability
scale.

\subsection{Uncertainty}

We report 95\% bootstrap intervals \cite{efron1993bootstrap} that jointly
resample profiles and seeds for retrieval, and participants and seeds for
probes and disease metrics, keeping arm and seed pairing intact. Participants,
not matched-pair blocks, are the resampling unit for disease metrics. We
interpret these pointwise intervals separately for each comparison. Locking
the external protocol removes analysis-selection freedom from the external
runs, but not from the UKCOVID discovery stage. For the external primary
contrast, we also resample at the matched-set level. CODA and Cambridge
resample the 100 matched case--control pairs and seeds, whereas Coswara
resamples exact matching strata and seeds. An interval that includes zero does
not establish equivalence. All AUROC and MRR values below are on a 0--100
scale, so AUROC contrasts are percentage-point differences.

\section{Results}
\label{sec:results}

\begin{table}[t]
\centering
\caption{UKCOVID paired contrasts. Point estimates with pointwise 95\%
intervals on the 0--100 scale, so AUROC contrasts are percentage-point
differences. Bold marks intervals that exclude zero. OPERA denotes OPERA-CT.}
\label{tab:contrasts}
\setlength{\tabcolsep}{1.6pt}
\scriptsize
\begin{tabular*}{\columnwidth}{@{}ll@{\extracolsep{\fill}}ccc@{}}
\toprule
Task & Contrast & AST-6L & OPERA & HeAR \\
\midrule
Sex probe & $C-W$ &
\textbf{19.1 [16.7,21.5]} &
\textbf{11.2 [9.4,13.1]} &
\textbf{8.6 [7.3,9.9]} \\
 & $C-\wys{}$ &
$-0.0$ [$-1.5$,1.5] &
0.7 [$-0.5$,1.9] &
0.6 [$-0.1$,1.4] \\
 & $\wys{}-W$ &
\textbf{19.2 [16.8,21.4]} &
\textbf{10.6 [9.0,12.2]} &
\textbf{8.0 [6.7,9.3]} \\
\addlinespace
Retrieval & $C-W$ &
\textbf{0.3 [0.2,0.5]} &
\textbf{0.3 [0.1,0.5]} &
\textbf{0.7 [0.5,1.0]} \\
 & $C-\wys{}$ &
0.2 [0.0,0.3] &
0.1 [$-0.1$,0.3] &
\textbf{0.4 [0.2,0.6]} \\
\addlinespace
Disease & $C-W$ &
0.6 [$-1.4$,2.8] &
$-0.2$ [$-1.9$,1.7] &
0.3 [$-1.1$,1.6] \\
 & $C-\wys{}$ &
$-0.0$ [$-2.3$,2.4] &
0.1 [$-1.7$,2.0] &
1.3 [$-0.2$,2.9] \\
 & $\wys{}-W$ &
0.7 [$-2.0$,3.1] &
$-0.3$ [$-2.2$,1.6] &
$-1.0$ [$-2.6$,0.5] \\
\bottomrule
\end{tabular*}
\end{table}

\begin{figure*}[t]
\centering
\includegraphics[width=0.80\textwidth]{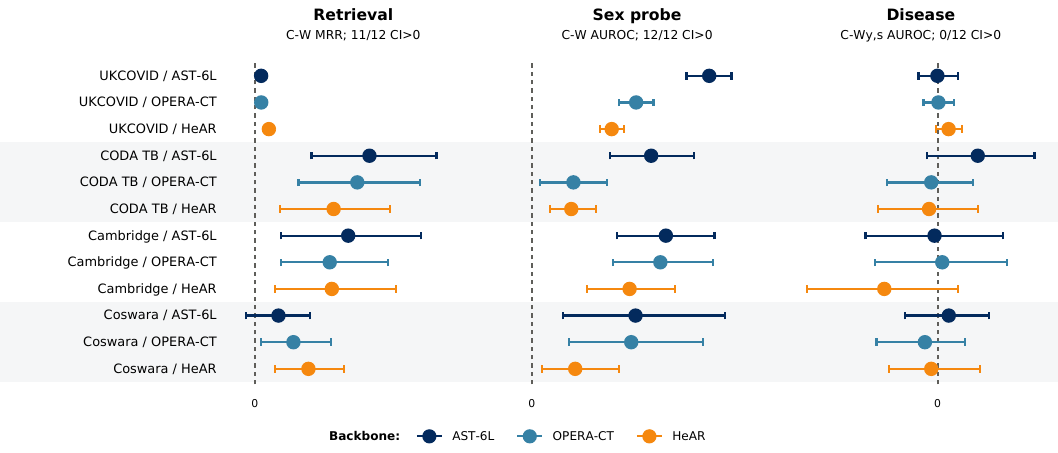}
\vspace{-5pt}
\caption{Contrasts on the discovery set and the external panel. Retrieval and
sex-probe panels show $C-W$; the disease panel shows the prespecified contrast
$C-\wys{}$, with matched-set or matching-stratum bootstrap for the external
datasets. Colors denote backbones (AST-6L, OPERA-CT, HeAR); x-axes are on the
0--100 scale. Full numeric results are in the supplementary logs.}
\label{fig:forest}
\end{figure*}

\vspace{-5pt}
\subsection{A label-preserving shuffle differs from a global shuffle}

The choice of shuffle matters even though neither $W$ nor $G$ preserves the
original participant pairing. For AST on UKCOVID, the source-split $W-G$ probe
AUROC differences are $6.46$ for COVID, $7.58$ for cough, and $6.47$ for
recruitment source, all with intervals excluding zero. On the matched split,
the COVID difference decreases to $1.26$ with an interval covering zero, while
the cough difference remains $8.03$. Restricting metadata donors by disease
label changes what a probe can decode without restoring any original
participant pair. A direct $C$-versus-$G$ comparison would therefore combine
this label-level effect with whatever exact pairing adds.

\subsection{Correct pairing changes retrieval and sex decodability}

Table~\ref{tab:contrasts} shows that changing only the metadata donor produces
a measurable difference in the frozen representation. Retrieval $C-W$ is
positive for all three backbones and across all five seeds for each backbone,
confirming that the pairing manipulation works. This is a necessary check, but
it does not establish complete profile recovery or greater clinical utility. For AST-6L, the sex-probe AUROC difference decreases from $19.1$ points under
$C-W$ to approximately zero under $C-\wys{}$. OPERA-CT and HeAR show the same
pattern, while $\wys{}-W$ remains large and positive. Much of the observed
$C-W$ difference can therefore also arise when the control preserves
recorded-sex consistency. The comparison does not isolate sex from correlated
metadata attributes, since \wys{} also preserves metadata correlated with sex
and a fraction of its pairs share identical profile strings. Raw audio is more
sex-decodable than any aligned representation, so a positive $C-W$ difference
is a change relative to the shuffled control, not a gain over the backbone.
Age probes improved under $C$ for AST but not OPERA. Because we did not build
an age-preserving control, age is not isolated.

Retrieval differs in one respect. The residual $C-\wys{}$ MRR difference is
positive for all three backbones, with intervals above zero for AST and HeAR,
although the AST lower bound is zero and the OPERA interval includes zero.
Preserving sex removes most of the sex-probe effect but not all of the
profile-correspondence effect. The residual persists when the candidate pool
is restricted by sex or by label and sex (supplementary logs).

\subsection{No consistent change in matched disease prediction}

Matched disease AUROC shows no consistent change. The $C-W$ difference changes
sign across backbones, and the correctly paired projector is not uniformly
better than raw audio, whose matched AUROC is $53.8$ for AST-6L, $57.7$ for
OPERA-CT, and $54.2$ for HeAR. The prespecified $C-\wys{}$ contrast is also
inconclusive, at $-0.03$ for AST, $0.09$ for OPERA, and $1.30$ for HeAR, with
all intervals covering zero (Table~\ref{tab:contrasts}). The HeAR interval
[$-0.23$, $2.87$] still allows a modest positive effect, so we would not call
this a null. $\wys{}-W$ is also mixed in sign. The disease differences are an
order of magnitude smaller than the sex-probe differences and are not
consistent in sign. The fixed MLP does not change this interpretation, and the target-assisted
diagnostics show no consistent cross-backbone advantage either (supplementary
logs).

Figure~\ref{fig:forest} shows the corresponding comparisons for the external
panel. Across the 12 dataset--backbone settings, every retrieval $C-W$ point
estimate is positive, and 11 of 12 intervals exclude zero. All 12 sex-probe
$C-W$ intervals exclude zero. For disease $C-\wys{}$, five estimates are
positive and seven are negative, and none of the intervals excludes zero. The
external targets are small and differ in disease and candidate library, so we
make no pooled claim. However, the pattern is the same on every dataset.
Retrieval and sex-probe contrasts are consistently positive, whereas matched
disease contrasts are not.

\subsection{Sensitivity analyses}

None of the sensitivity analyses changes this interpretation. Adding raw audio
to metadata as a reference does not give the aligned representations a
consistent advantage over metadata plus raw audio. Source-calibrated NLL
differences decrease after target recalibration while AUROC remains unchanged,
pointing to probability-scale miscalibration rather than a difference in
ranking.

\section{Discussion and Limitations}
\label{sec:discussion}

\textbf{What shifts the representation.} The results do not establish that
audio--metadata alignment fails. Correct pairing improves profile retrieval
and makes attributes more decodable. However, we find no consistent
improvement in matched disease prediction, and the intervals do not rule out
modest benefits or harms. The $C/W/G/\wys{}$ hierarchy shows why a single
ablation cannot distinguish these pairing effects. The label mix, recorded
sex, and exact participant identity each shift the representation on their
own, so a two-arm correct-versus-shuffled comparison attributes all three to
``alignment.'' Disease prediction needs to be measured separately from
alignment quality, against raw-audio references.

\textbf{What this means for reporting.} Reports of alignment gains should
specify which outcome improved. We suggest reporting profile correspondence,
attribute decodability, and matched disease prediction separately, each with a
raw-audio reference and explicit uncertainty. We also suggest using a
label-preserving control rather than a global shuffle whenever the metadata
are correlated with the outcome.

\textbf{Limitations.} The external targets have limited precision, and
matching balances measured covariates only, so unmeasured confounding remains.
Pointwise bootstrap intervals do not account for choices made during the
UKCOVID discovery stage. Participant separation within a study does not rule
out exposure through upstream pretraining, as OPERA-CT illustrates. The
sex-preserving control isolates neither age nor other attributes and may
retain structure correlated with its conditioning variables. We study one
projector design and one deterministic schema rendering, not free-form
prompts, other text encoders, or full multimodal downstream systems. A
synthetic study is included as supplementary material but is not relied on,
because it failed its preregistered gates for a general mechanism, so we make
no such claim. A disease-invariant variant of the objective was planned but
could not be trained because the cohorts lacked the required cross-environment
same-label positives and within-environment opposite-label negatives. This is
a feasibility limit, not a negative result for the method.
\vspace{-5pt}

\section{Conclusion}

Across the discovery dataset and an unchanged external rerun, contrastive
audio--metadata alignment improves retrieval of the paired profile relative to
within-label shuffling and, in most comparisons, relative to sex-preserving
shuffling. We find no consistent evidence that this improvement in profile
correspondence leads to better disease prediction on matched test sets.
Demographic structure and residual profile correspondence persist under the
same pipeline. Reports of clinical audio alignment should therefore
distinguish correspondence from disease transfer.
\section*{\small Compliance with Ethical Standards and Acknowledgment}
This work is a retrospective analysis of de-identified datasets accessed and
used under their respective data-use terms. GPT-5.5 and
Claude were used to assist with language polishing.
\bibliographystyle{IEEEbib}
\bibliography{references}

\end{document}